\documentclass{article}

\usepackage{microtype}
\usepackage{graphicx}
\usepackage{subfigure}
\usepackage{booktabs} 

\usepackage{hyperref}

\usepackage[accepted]{mlsys2024}

\usepackage{color}
\usepackage{listings}
\usepackage{multicol}
\usepackage{multirow}
\usepackage{xcolor}
\usepackage{amssymb}
\usepackage{amsmath}

\lstdefinelanguage{JavaScript}{
  morekeywords=[1]{break, continue, delete, else, function, if, in,
    new, this, typeof, var, void, while, with, constructor},
  morekeywords=[2]{false, null, true, boolean, number, undefined,
    Array, Boolean, Date, Math, Number, String, Object},
  morekeywords=[3]{eval, parseInt, parseFloat, escape, unescape},
  sensitive,
  morecomment=[s]{/*}{*/},
  morecomment=[l]{//},
  morecomment=[s]{/**}{*/}, 
  morestring=[b]{'},
  morestring=[b]{"},
}[keywords, comments, strings]

\lstalias[]{ES6}[ECMAScript2015]{JavaScript}
\lstdefinelanguage[ECMAScript2015]{JavaScript}[]{JavaScript}{
  morekeywords=[1]{async, case, catch, class, const, default, do,
    enum, export, extends, finally, from, implements, import, instanceof,
    let, static, super, switch, throw, try},
  morekeywords=[4]{await, return, for},
  morestring=[b]{`} 
}

\lstdefinelanguage[distmljsSample]{JavaScript}[ECMAScript2015]{JavaScript}{
  morekeywords=[2]{MLPModel, K, nn, core, Layer, layers, Promise, VariableResolvable, functions, images, labels, trainLoader, optimizer},
  morekeywords=[5]{c, relu, Linear, tidy, Variable, to, softmaxCrossEntropy, zeroGrad, backward, step},
}

\definecolor{basecolor}{HTML}{081884}
\definecolor{identifiercolor}{HTML}{081884}
\definecolor{commentcolor}{HTML}{028002}
\definecolor{strcolor}{HTML}{a6201f}
\definecolor{keywordcolor1}{HTML}{1010ff}
\definecolor{keywordcolor2}{HTML}{2d839e}
\definecolor{keywordcolor4}{HTML}{af01db}
\definecolor{keywordcolor5}{HTML}{795e27}

\mlsystitlerunning{WgPy: GPU-accelerated NumPy-like array library for web browsers}

\begin{document}

\twocolumn[
\mlsystitle{WgPy: GPU-accelerated NumPy-like array library for web browsers}



\mlsyssetsymbol{equal}{*}

\begin{mlsysauthorlist}
\mlsysauthor{Masatoshi Hidaka}{ut}
\mlsysauthor{Tatsuya Harada}{ut,riken}
\end{mlsysauthorlist}

\mlsysaffiliation{ut}{The University of Tokyo, Japan}
\mlsysaffiliation{riken}{RIKEN, Japan}

\mlsyscorrespondingauthor{Masatoshi Hidaka}{hidaka@mi.t.u-tokyo.ac.jp}

\mlsyskeywords{Machine Learning, MLSys, Web, GPU, WebGL, WebGPU}

\vskip 0.3in

\begin{abstract}
To execute scientific computing programs such as deep learning at high speed, GPU acceleration is a powerful option. With the recent advancements in web technologies, interfaces like WebGL and WebGPU, which utilize GPUs on the client side of web applications, have become available. On the other hand, Pyodide, a Python runtime that operates on web browsers, allows web applications to be written in Python, but it can only utilize the CPU, leaving room for acceleration. Our proposed new library, WgPy, provides array computation capabilities on the GPU with a NumPy-compatible interface in the web browser. This library not only implements array operations such as matrix multiplication on WebGL and WebGPU, but also allows the users to write custom kernels that can run on GPUs with minimal syntax knowledge, allowing you to run a variety of algorithms with minimal overhead. WgPy also implements a special thread synchronization mechanism, which bridges asynchronous semantics of JavaScript with Python's synchronous semantics, allows code written for CuPy, the NumPy-compatible array library for CUDA, to run directly in a web browser. In experiments involving training a CNN model, it achieved processing at 95 times the speed compared to CPU execution.
\end{abstract}
]



\printAffiliationsAndNotice{}  

\section{Introduction}
\label{submission}

NumPy~\cite{harris2020array} is a popular Python library that provides multi-dimensional numerical arrays and has become a de facto standard component for implementing scientific and technical computations, including machine learning. NumPy is used not only in machine learning libraries like scikit-learn and deep learning libraries like Chainer~\cite{chainer_learningsys2015} but also in data analysis libraries such as pandas, resulting in a vast software ecosystem that leverages NumPy.

While scientific computations are usually performed on servers or workstations, consumer devices like smartphones are also becoming widely adopted and increasingly powerful. Establishing an environment where scientific computations can be executed on these devices has potential applications, such as processing data that requires privacy considerations (e.g., medical images) or performing large-scale computations through volunteer computing by coordinating multiple devices for distributed computing. Notably, web applications offer the advantage of being accessible by simply opening a URL, making them easy to use regardless of the user's IT literacy.

JavaScript is the primary implementation language for web applications, but the lack of a de facto standard array library and the fact that JavaScript syntax is less intuitive than Python can be implementation hurdles. For example, when adding array objects a and b, Python allows the intuitive expression a + b using operator overloading. In contrast, JavaScript does not support operator overloading, necessitating less intuitive expressions like a.add(b). Compared to Python environments, there are significantly fewer scientific and technical computing libraries available in JavaScript environments. If existing assets based on the Python and NumPy combination could be utilized in web applications, it would substantially simplify the implementation of web applications for scientific computations.

Recently, web browsers have been able to execute not only JavaScript but also a new type of code called WebAssembly. Algorithms implemented in languages such as C or Rust can be compiled into WebAssembly for execution. Applying this technology, Pyodide~\cite{pyodide_2021} compiles the Python interpreter into WebAssembly, making it possible to run Python in the web browser. Since NumPy is available on Pyodide, scientific computing code can be executed directly. However, NumPy running on Pyodide can only use a single-threaded CPU, limiting its computational speed.

It is known that processing multi-dimensional arrays, especially matrix multiplication, can be significantly accelerated by using a GPU. On native platforms where NVIDIA GPUs (CUDA) are available, CuPy~\cite{cupy_learningsys2017} provides NumPy-compatible multi-dimensional array processing and enables fast computations. CuPy was developed to make the deep learning library Chainer GPU-compatible, but has also been applied to other fields such as natural language processing~\cite{honnibal2020spacy} and healthcare imaging~\cite{monai2022}. CuPy only works on NVIDIA GPUs as it uses CUDA, but ClPy~\cite{clpy2019}, which uses OpenCL, a more general standard for controlling compute accelerators, is also developed, enabling NumPy-compatible array processing on various GPUs. However, CuPy and ClPy do not work in the Pyodide environment, which does not have access to CUDA or OpenCL, and therefore cannot be utilized for accelerating web applications. The graphics APIs built into web browsers to control GPUs include standards like WebGL and WebGPU, necessitating array libraries compatible with these standards.

In this study, we propose a new library, WgPy, which wraps the graphics APIs specific to web browsers and enables array processing using the GPU with a NumPy-compatible interface. As an application example, we also conduct experiments on deep learning and optimizing hyperparameters for deep learning through distributed computing.

The contributions of this study are as follows:

\begin{itemize}
\item By providing an array library that allows the use of the GPU from Python code running in the web browser, we support the development of fast and installation-free scientific computing applications.
\item We construct the necessary techniques for implementing a NumPy-compatible array computation library on the WebGL and WebGPU standards and establish methods to apply these standards to scientific computations.
\end{itemize}

We have released the developed WgPy as open-source software \footnote{\url{https://github.com/mil-tokyo/wgpy}}.

\section{Related works}
\subsection{Array Processing}
Accelerated with a NumPy interface CuPy is an array processing library with a NumPy-like interface that enables fast array computations using a GPU via the CUDA environment. The contents of the arrays are stored in the GPU memory, and computations are executed using CUDA kernels. In addition to implementing kernels for basic operations such as matrix indexing and matrix multiplication, application developers can also implement custom kernels. A typical implementation method is ``ElementwiseKernel''. An example of ElementwiseKernel is shown source code \ref{elementwise-cupy}. In this case, the squared diff is computed for each element of the two arrays. Using ElementwiseKernel, the two operations can be combined in a register or cache operation. The use of the ElementwiseKernel makes it possible to combine the two operations into a single operation on a register or in the cache. Using a unique kernel in this way makes it possible to efficiently process tasks that cannot be efficiently implemented by combining basic operations. The kernel description needs to follow the syntax of CUDA C++, but the processes such as GPU initialization and kernel compilation are hidden within the library, thus keeping the learning cost of CUDA-specific knowledge low.

\begin{figure}[t]
  \begin{lstlisting}[caption=Example element-wise kernel of CuPy,label=elementwise-cupy]
ElementwiseKernel(
  in_params='float32 x, float32 y',
  out_params='float32 z',
  operation='z = (x - y) * (x - y)',
  name='squared_diff'
)
    \end{lstlisting}
\end{figure}

ClPy is an array processing library implemented with reference to CuPy, but it uses the OpenCL environment instead of CUDA. The OpenCL environment supports a wide range of accelerators, including AMD GPUs. To execute code written for CuPy without modification, ClPy includes a mechanism that parses kernels written in CUDA C++ and converts them into OpenCL C code, which can be compiled in the OpenCL environment.

In this research, we develop a library that provides functionality equivalent to CuPy and ClPy in a web browser environment where CUDA and OpenCL cannot be used.

Spartan is an array processing library that supports distributed computation with a NumPy-like interface. This library is designed for low-latency communication between arbitrary nodes and is not directly applicable to communication between web browsers. Distributed computation will be discussed at the upper application layer in the experimental section.

\subsection{Array Processing in Web Browsers}
GPU.js~\cite{gpujs2018} is a library that parses array processing functions written in JavaScript, converts them into GLSL code for WebGL, and makes them executable on the GPU. Since all array processing needs to be written in JavaScript, it is not possible to utilize code assets implemented using NumPy.

TensorFlow.js~\cite{smilkov-2019} is a JavaScript library for web browsers with an interface similar to the deep learning library TensorFlow. It can load and execute models trained in TensorFlow, providing excellent interoperability at the level of deep learning.

In this research, we develop an array processing library that runs in a Python environment within a web browser, making it convenient to utilize existing code assets and knowledge that use Python. Furthermore, by targeting array processing that is positioned at a lower layer than deep learning, this approach is expected to be applicable to a wide range of tasks.

\section{Implementation}
First, we introduce GPU interface APIs available for web browsers.

\subsubsection{WebGL}
WebGL, which has been available since 2011, is currently the most popular GPU interface. With WebGL, it became possible to render 3D graphics using the GPU on a web browser. WebGL allows users to describe processes executed on the GPU using a shader language called GLSL ES. However, WebGL is specialized for graphics use and does not have functionalities for general-purpose scientific computations. One method to perform scientific computations using this graphics-oriented interface is to utilize fragment shaders. Fragment shaders calculate the color of each pixel when a 3D object is displayed on the screen. Textures of objects can be used as input. Instead of outputting the calculation results of fragment shaders to the screen, they can be written to a texture using a mechanism called a framebuffer. Although textures represent images, they can be used as general-purpose numerical arrays since arbitrary values (integers or floating-point numbers) can be written to each pixel \footnote{Our implementation assumes WebGL version 2.0. In version 1, there were additional restrictions, such as only supporting 8-bit integers for framebuffer output.}

However, since texture is inherently designed for handling images, various constraints arise. (1) Pixels in a texture are represented by a two-dimensional index, and the maximum length of one side is 4096 (device-dependent). Arrays that do not fit within the maximum texture size can be represented using a mechanism called texture arrays, but the implementation becomes complex. (2) The processing of each output pixel is independent. In cases like matrix multiplication, where multiple output pixels share some of the input, there are efficient memory access methods available in other GPU interfaces, but they cannot be used in WebGL.

\subsubsection{WebGPU}
WebGPU is a new GPU interface that began implementation around 2017. After various discussions about the shader language, a new language called WGSL was developed, and an implementation capable of executing WGSL is available in Google Chrome. For scientific computations, the important aspect of WGSL is its implementation of compute shaders. Compute shaders are specialized for numerical computations, eliminating the constraints that arise when applying graphics-oriented functionalities of WebGL to scientific computing. As of 2024, smartphones capable of running WebGPU will be limited to a few Android-based models, so it will be necessary to use WebGPU in combination with WebGL to support a wide variety of devices. In WgPy, WebGL or WebGPU is automatically selected according to the GPU interface supported by the device unless the user writes a custom kernel with shader source code.

\subsubsection{Implementation of a Numpy-like Interface}

\begin{figure}[t]
  \centering
  \includegraphics[width=0.8\linewidth]{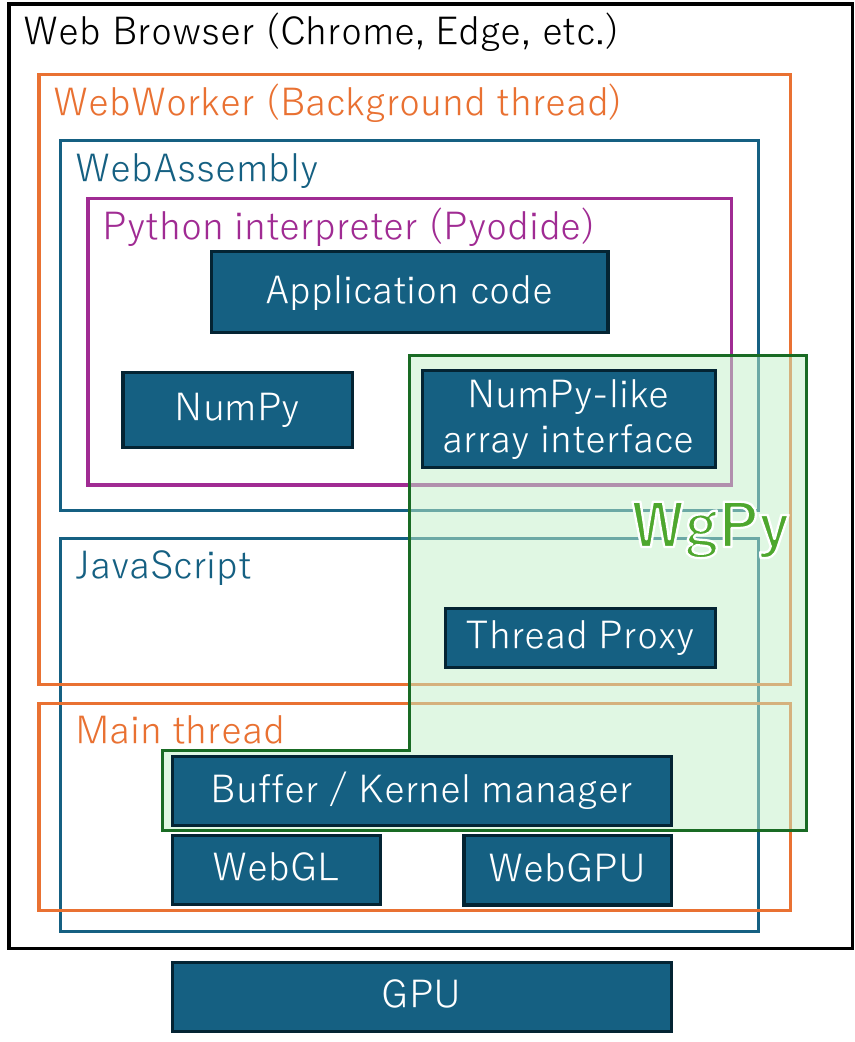}
  \caption{The structure of WgPy. WgPy exposes NumPy-like array interface in the Python interpreter, and intermediates with array processing routines implemented in WebGL and WebGPU, the GPU interfaces for web browsers.}
  \label{fig:system}
\end{figure}

\begin{figure*}[t]
\begin{minipage}{.50\textwidth}
  \begin{lstlisting}
# Step 1: Run calculations on GPU
gpuArray = gpuArrayA + gpuArrayB
# Step 2: call asnumpy() function to copy result to CPU
cpuArray = asnumpy(gpuArray)
# Step 7: access result on CPU.
# However, actual data copy is not yet complete.
print(cpuArray[0])

def asnumpy(gpuArray) {
  # Buffer to be filled by JavaScript
  cpuArray = np.zeros(gpuArray.shape)
  # Step 3: call JavaScript function to copy result to CPU
  toCPU(gpuArray.bufferId, cpuArray)
  # Step 6: resume execution of Python code
  return cpuArray
}
    \end{lstlisting}
\end{minipage}
\begin{minipage}{.50\textwidth}
  \begin{lstlisting}
function toCPU(gpuBufferId, bufferInsidePython) {
  const buffer: GPUBuffer = buffers[gpuBufferId];
  // Step 4: Start GPU to CPU data transfer.
  buffer.mapAsync(GPUMapMode.READ).then(() => {
    // Step 8: This callback function is called when transfer is ready. Copying GPU to CPU can be implemented here.
  }
  // Step 5: mapAsync() returns immediately. As a result, toCPU() returns.
}
    \end{lstlisting}
\end{minipage}
  \caption{Pseudo code to transfer data from GPU to CPU, which \textbf{does not work}. Left: Python side, Right: JavaScript side.}
  \label{fig:transfer-cpu-gpu}
\end{figure*}
In CuPy, not only can operations between arrays on the GPU be implemented with syntax equivalent to NumPy, but it also provides an interface for transferring NumPy arrays on the CPU to the GPU and vice versa. In this study, we implement an interface for CPU-GPU array transfer similar to CuPy. This allows replacing the backend of libraries like Chainer that depend on CuPy with WgPy, enabling them to run in a web browser.

The structure of the system is shown in Fig. \ref{fig:system}. WebGL and WebGPU can only be called from JavaScript; they cannot be called directly from the Python processing system, which is built and runs on WebAssembly. Therefore, from Python, call a function implemented in JavaScript using the Pyodide functionality, and call the GPU interface within that function.
In the web browser, there is a problem that if a high-load calculation is performed on the main thread, the user interface cannot be processed and appears to be frozen; when a high-load calculation is performed on the CPU, it is processed on a separate thread using a mechanism called WebWorker. On the other hand, WebGL and WebGPU can only be called from the main thread. Therefore, the Python processing system is run on the WebWorker, and communication with the main thread is done by JavaScript.
In WebGL, arrays are stored in textures; in WebGPU, arrays are stored in buffers. CuPy allows a large buffer (memory pool) to be allocated and divided into a set of small array variables, reducing the overhead of memory allocation. However, the current WebGPU implementation does not implement this because allocating a large buffer (several hundred MB) at once causes an error.

\subsubsection{Synchronous Use of Asynchronous JavaScript API from Python}

\begin{figure*}[t]
\centering
  \includegraphics[width=1\linewidth]{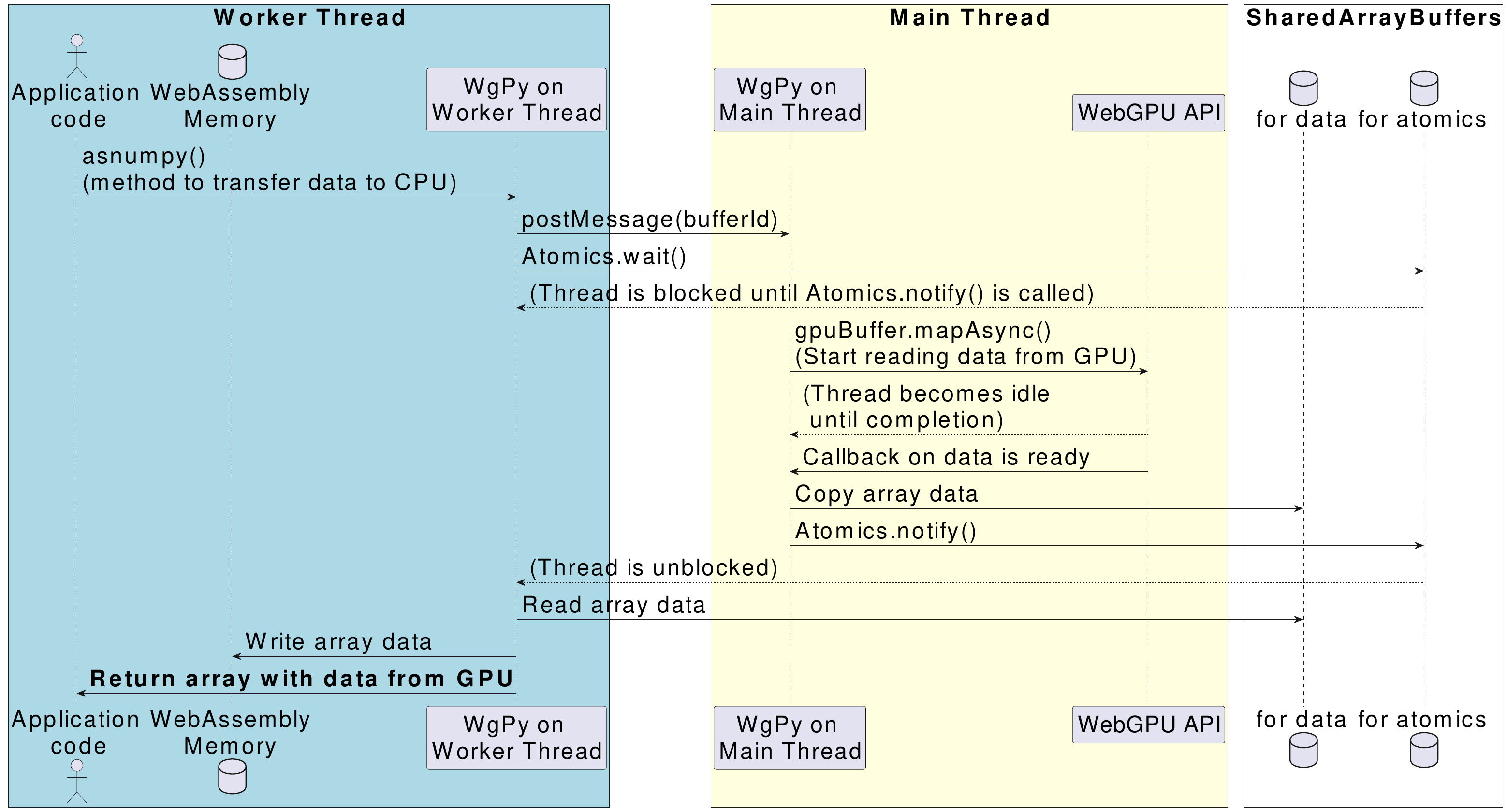}
  \caption{Sequence diagram showing the process of transferring array data on the GPU to a NumPy array on the CPU, using the Atomics API to block worker threads, so that Python code does not need to be aware of asynchronous processing and can run without modifying existing code that uses CuPy.}
  \label{fig:atomic-sequence}
\end{figure*}

Asynchronous operation is common in JavaScript. WebGPU requires asynchronous operations using \verb|Promise| in data transfer between CPU and GPU. On the other hand, Python code using NumPy / CuPy assumes synchronous processing, and it is usually difficult to intervene asynchronous processing when obtaining the result of array processing. This issue is explained in pseudo code \ref{fig:transfer-cpu-gpu}.
WgPy aims to make the source code that uses CuPy executable without modification. In CuPy, the \verb|asnumpy| function is used to transfer an array on the GPU to a NumPy array on the CPU. Immediately after calling this function, a NumPy array with a copy of the data on the GPU must be obtained. However, in WgPy with WebGPU backend, the \verb|mapAsync| function for copying from the GPU to the CPU returns immediately upon starting the process, and the actual copying is done asynchronously. Therefore, it is necessary to work around the \verb|asnumpy| function to make the result available immediately after it returns.

To address this problem, a synchronization mechanism using SharedArrayBuffer and Atomics API was implemented. The process steps are shown in Fig \ref{fig:atomic-sequence}. SharedArrayBuffer plays two roles: sharing the contents of the array between threads and using the Atomics API to wait for WebWorker threads. As mentioned above, Python code runs on the WebWorker thread, while the WebGPU API can only operate on the main thread. The mechanism works as follows: (0) SharedArrayBuffer, which can share numeric values between the threads are allocated in advance. (1) The application code requests WgPy to transfer GPU array to NumPy Array by \verb|asnumpy| method. (2) WgPy code on the worker thread retrieves the GPU buffer id of the array and send it to the main thread using \verb|postMessage| API. (3) The worker thread calls \verb|Atomics.wait| of Atomics API. It takes a SharedArrayBuffer as the argument. By calling the method, the caller thread is blocked (sleeps) until other thread calls \verb|Atomics.notify| with the same SharedArrayBuffer. (4) The main thread starts the data transfer from the specified GPU buffer to the CPU using \verb|GPUBuffer.mapAsync|. The function call immediately returns and the main thread become idle. (5) When the GPU data is ready to be accessed from CPU, the callback function is called. The main thread copies the contents of the array to the SharedArrayBuffer. (6) The main thread calls \verb|Atomics.notify| to resume the worker thread. (7) The worker thread resumes and copies the contents from the SharedArrayBuffer to the area of the web assembly's memory reserved for the resulting NumPy array. (8) The \verb|asnumpy| method returns with the NumPy array with the data from GPU. To the user, it appears that data transfer has been completed synchronously.

\subsubsection{Workarounds for CuPy Kernels}
In CuPy, it is possible to implement custom kernels using the CUDA language. For example, this is used to implement the backward function of ReLU. When using NumPy functions, it can be implemented as \verb|np.where(y > 0, gy, 0)|, which requires two operation calls (\verb|y > 0| and \verb|np.where|). By implementing this with a custom kernel, it will be more efficient. With ElementwiseKernel, processing can be applied to all elements of an array by describing element-wise operations without knowing the entire API of CUDA. A similar function was implemented in WgPy. The ReLU backward function equivalent to CuPy is shown in the source code \ref{relu-backward-webgl} and \ref{relu-backward-webgpu}. Since the shader languages differ for WebGL and WebGPU, the user needs to implement different code. It is also possible to implement loops and refer to multiple input elements, but in WebGL, the number of loop iterations must be constant, so it is necessary to embed numerical values into the code through string processing. In Chainer, tens of custom kernels are used, including the backward of ReLU, making automatic conversion difficult. Therefore, they were manually translated into WebGL/WebGPU and implemented in the library.

\begin{figure}[t]
  \begin{lstlisting}[caption=ReLU backward custom kernel in WebGL,label=relu-backward-webgl]
ElementwiseKernel(
  in_params="float y, float gy",
  out_params="float gx",
  operation="gx = y > 0.0 ? gy : 0.0",
  name="relu_bwd",
)
    \end{lstlisting}
\end{figure}

\begin{figure}[t]
  \begin{lstlisting}[caption=ReLU backward custom kernel in WebGPU,label=relu-backward-webgpu]
ElementwiseKernel(
  in_params="f32 y, f32 gy",
  out_params="f32 gx",
  operation="if (y > 0.0) { gx = gy; } else { gx = 0.0; }",
  name="relu_bwd",
)
    \end{lstlisting}
\end{figure}

\section{Experiments}
In this section, we present the benchmarks for array processing using WgPy. As shown in Table \ref{tab:devices}, five hardware and software combinations were used in the experiment to demonstrate that WgPy can work in a cross-platform environment. Currently, only Chrome supports WebGPU.

\begin{table}[t]
  \centering
  \caption{Device configurations used in the experiments.}
  \begin{tabular}{lll}
    \toprule
    OS & Hardware & Browser \\
    \midrule
    Windows & \begin{tabular}{l}AMD Ryzen7 5700X,\\NVIDIA RTX 4070\end{tabular} & Chrome \\
    macOS & MacBook Air 2020 (M1) & Chrome \\
    macOS & MacBook Air 2020 (M1) & Safari \\
    Android & Pixel 8 & Chrome \\
    iOS & iPhone 13 Mini & Safari \\
    \bottomrule
  \end{tabular}
  \label{tab:devices}
\end{table}

\begin{figure}[t]
  \centering
  \includegraphics[width=0.8\linewidth]{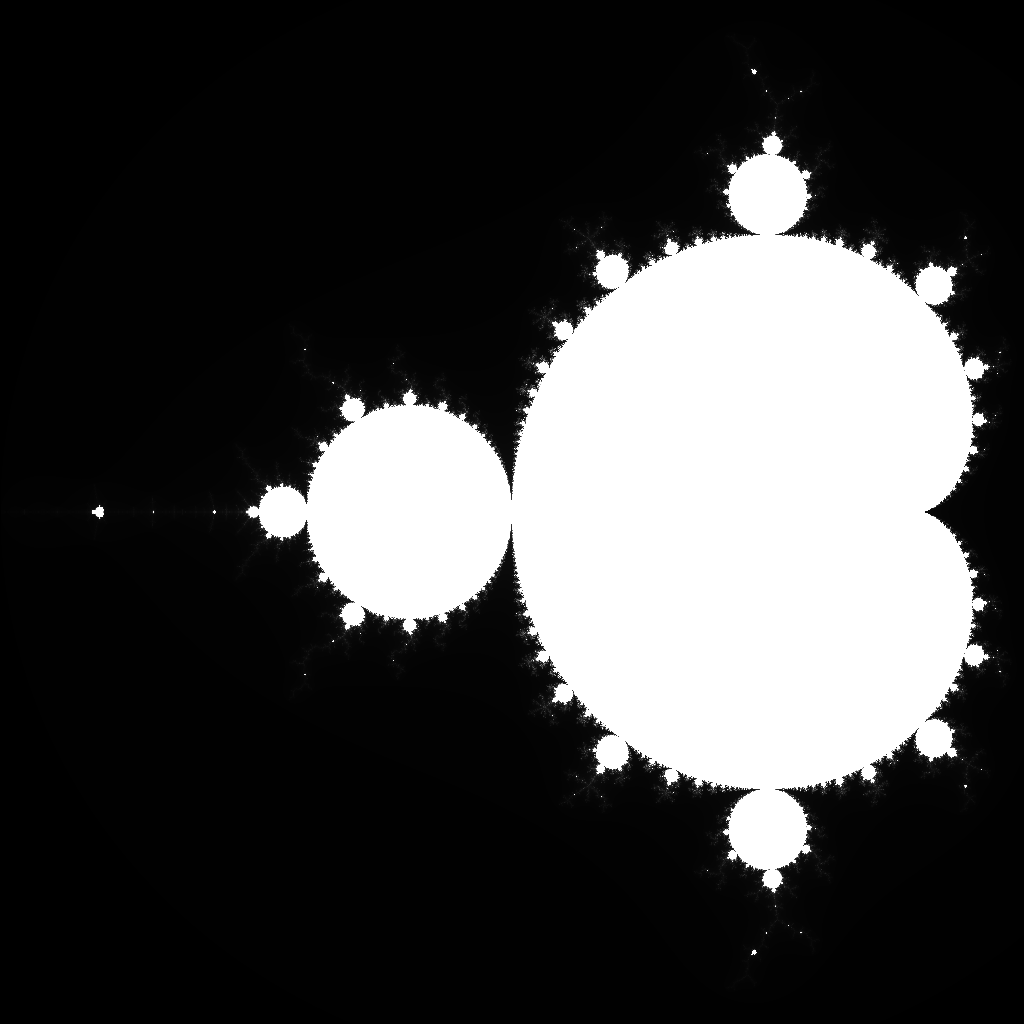}
  \caption{Visualization of Mandelbrot set. The range of the real axis is [-2.0, 0.5] and the range of the imaginary axis is [-1.2, 1.2]. White pixels indicate that the sequence does not diverge, and black pixels indicate that the sequence does diverge.}
  \label{fig:mandelbrot_visualize}
\end{figure}

\begin{figure}[t]
  \centering
  \includegraphics[width=1\linewidth]{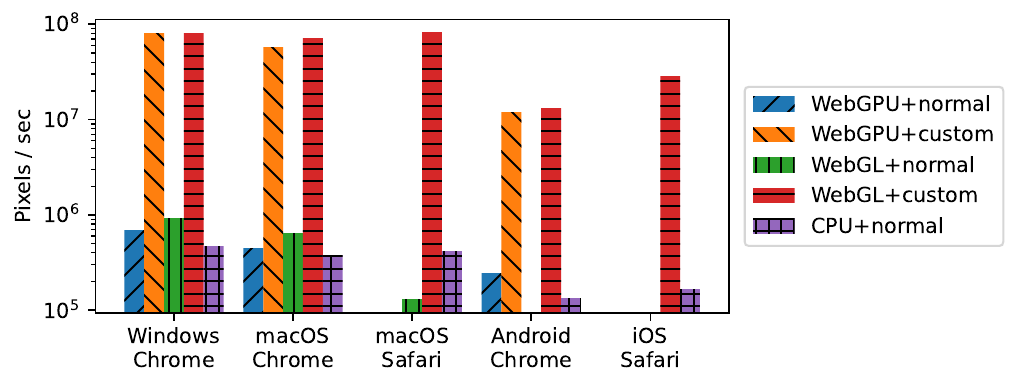}
  \caption{The speed of computing the Mandelbrot set. Normal indicates the case where the kernel is implemented using a combination of basic operations, and custom indicates the case where a custom kernel is used.}
  \label{fig:mandelbrot}
\end{figure}

\begin{figure*}[t]
  \begin{minipage}[b]{0.24\linewidth}
    \centering
    \includegraphics[scale=0.25]{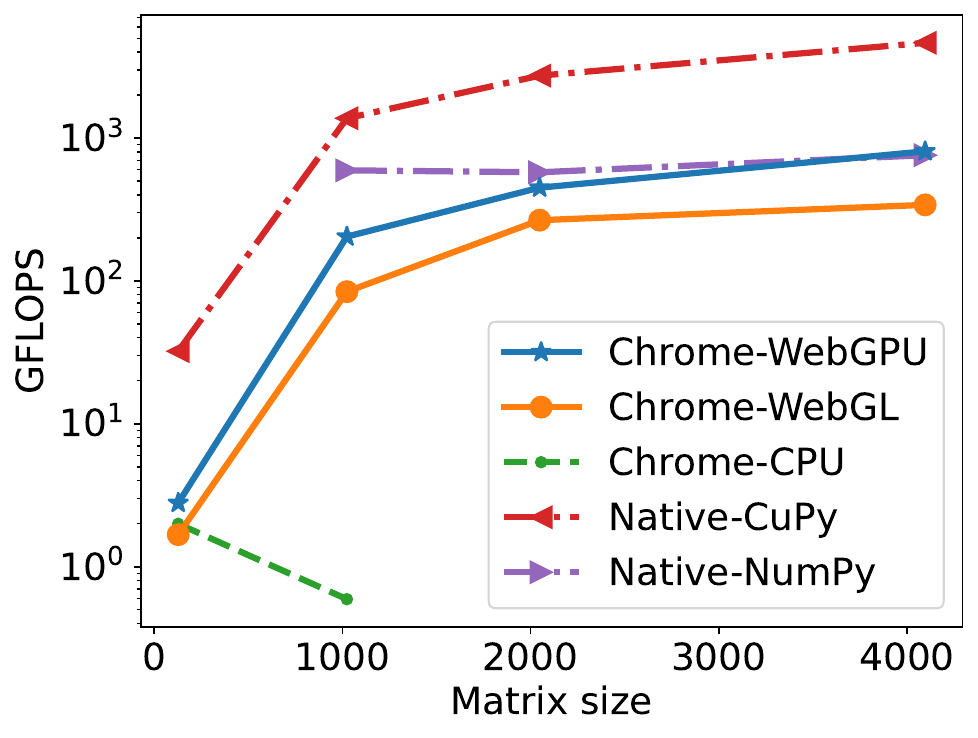}
  \end{minipage}
  \begin{minipage}[b]{0.24\linewidth}
    \centering
    \includegraphics[scale=0.25]{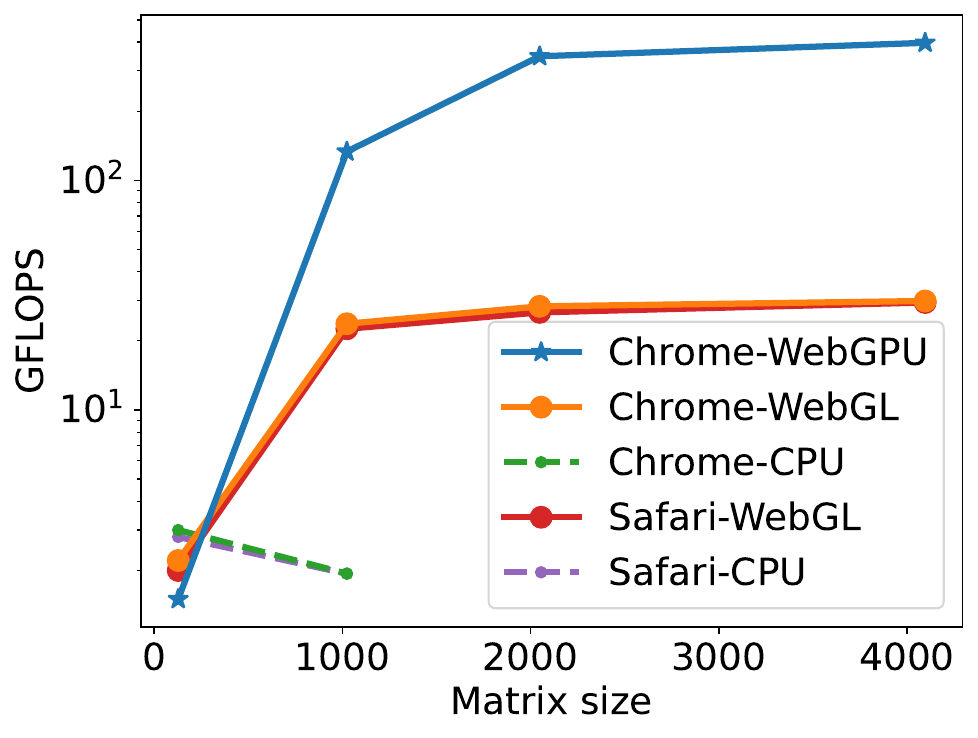}
  \end{minipage}
  \begin{minipage}[b]{0.24\linewidth}
    \centering
    \includegraphics[scale=0.25]{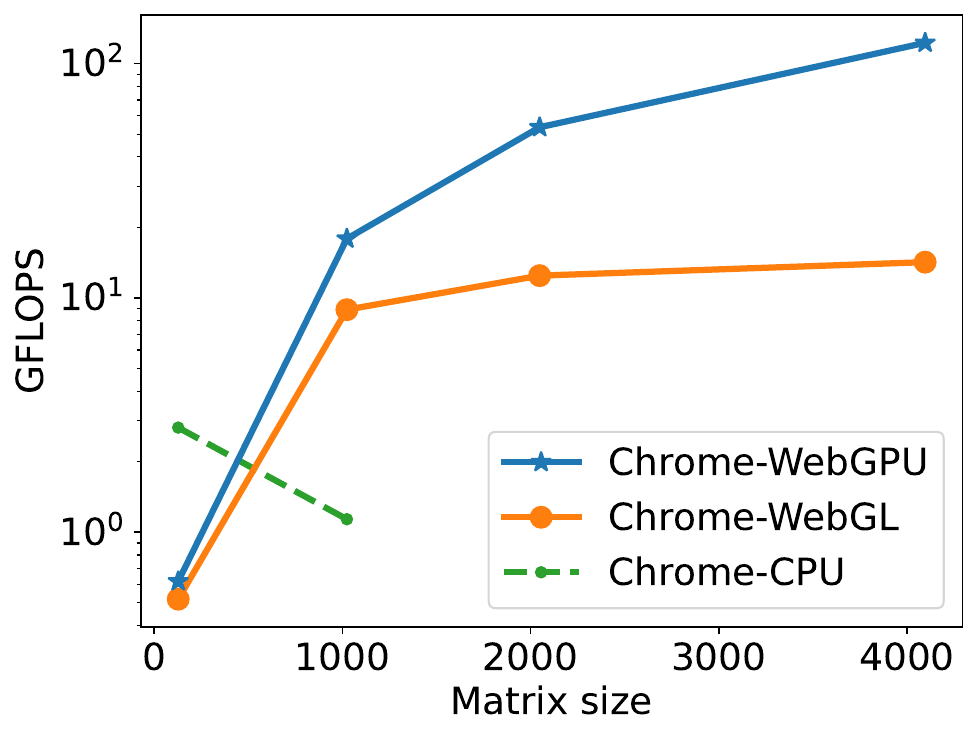}
  \end{minipage}
  \begin{minipage}[b]{0.24\linewidth}
    \centering
    \includegraphics[scale=0.25]{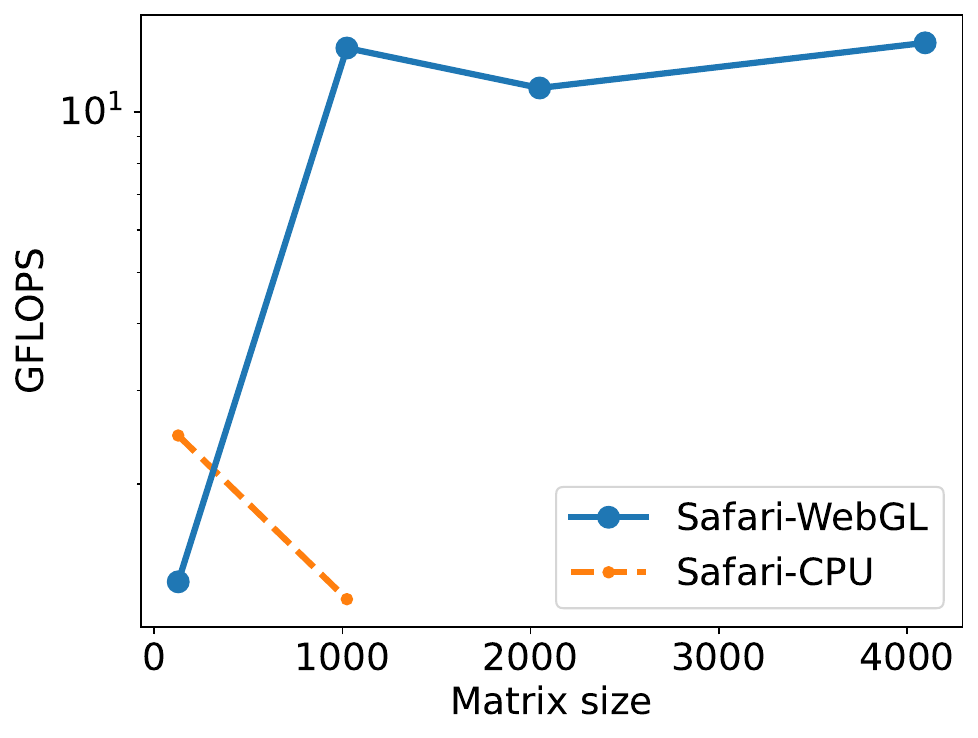}
  \end{minipage}
    \caption{The speed of matrix multiplication of two $N \times N$ square matrices. Device configurations are Windows, macOS, Android, iOS from left to right.}
    \label{fig:matmul}
\end{figure*}

\begin{figure}[t]
  \centering
  \includegraphics[width=1\linewidth]{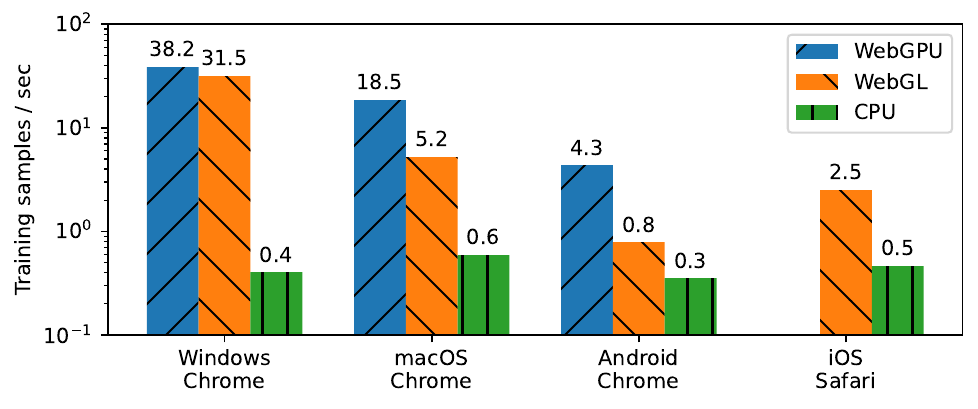}
  \caption{The training speed of ResNet-18 on different backends and device configurations.}
  \label{fig:resnet}
\end{figure}

\begin{figure}[t]
  \centering
  \includegraphics[width=0.8\linewidth]{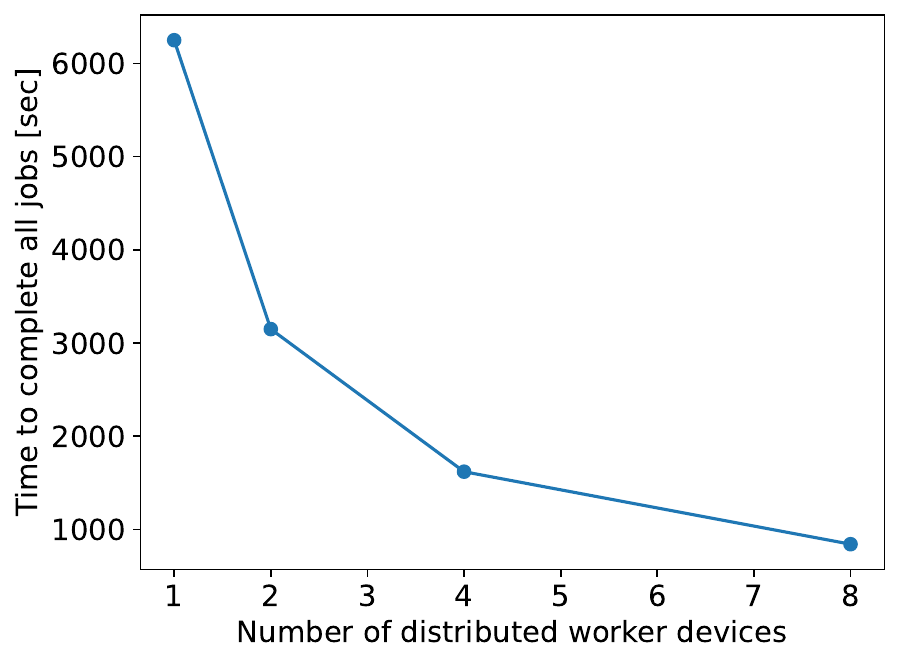}
  \caption{The processing time of naive neural architecture search with distributed workers}
  \label{fig:nas}
\end{figure}

\subsection{Mandelbrot set}
The Mandelbrot set is the set of complex numbers $c$ such that the following recurrence relation do not diverge.

\begin{equation}
  \left\{ 
  \begin{aligned}
    z_{n+1} & =  z_n^2 + c \\   
    z_0 & = 0& 
  \end{aligned} 
  \right.
\end{equation}

In this experiment, we benchmark the element-wise operation of WgPy by iteratively computing the $z_{500}$ for each $c$ sampled equally spaced from a grid of $1024 \times 1024$.
A straightforward implementation is shown in the source code \ref{mandelbrot-simple}. In this code, the ``\verb|cupy.get_array_module|'' method is used so that the same code can process both NumPy arrays and WgPy arrays. This method returns \verb|cupy| module if the argument array is on the GPU, otherwise returns \verb|numpy|. The keyword \verb|cupy| is used instead of \verb|wgpy|, in order to run the code implemented for CuPy without modification. Note that WgPy currently does not implement complex numbers, so the input is given as a pair of real numbers. In this implementation, individual GPU kernels are called for each operation, such as addition and multiplication, leading to a relatively high overhead for each call. Next, we show an implementation where processing is integrated into a single kernel using a custom kernel in the source code \ref{mandelbrot-webgpu}. An example of the generated image is shown in the Fig. \ref{fig:mandelbrot_visualize}. As a benchmark experiment, the time to process $1024 \times 1024$ complex numbers (pixels) was measured the backends and with and without custom kernel. The results are shown in Fig. \ref{fig:mandelbrot}. With a custom kernel, the speedup was 170 times faster than the CPU in the Windows and Chrome cases.
Without a custom kernel, the speed improvement from CPU was minor due to the large overhead of GPU calls. In cases where the processing time is long, the mobile device sometimes stopped processing. In order to deal with such situations, heuristics that split the processing and intersperse it with appropriate sleep are considered necessary.

\begin{figure}[t]
  \begin{lstlisting}[caption=Mandelbrot set implementation for NumPy / WgPy,label=mandelbrot-simple]
def mandelbrot(real, imag):
    xp = cupy.get_array_module(real)
    xs = xp.zeros((1024, 1024), dtype=np.float32)
    ys = xp.zeros((1024, 1024), dtype=np.float32)
    count = xp.zeros((1024, 1024), dtype=np.int32)
    for _ in range(500):
        # z = z * z.conj() + c
        xs, ys = xs * xs - ys * ys + real, xs * ys * 2.0 + imag
        count += ((xs * xs + ys * ys) < 4.0).astype(np.int32)
    return count
  \end{lstlisting}
\end{figure}

\begin{figure}[t]
  \begin{lstlisting}[caption=Custom WebGPU kernel of Mandelbrot set,label=mandelbrot-webgpu]
ElementwiseKernel(
    in_params="f32 real, f32 imag",
    out_params="i32 c",
    operation="""
c = 0;
var x: f32 = 0.0;
var y: f32 = 0.0;
for(var k: u32 = 0u; k < 500u; k = k + 1u) {
    var nx: f32 = x * x - y * y + real;
    var ny: f32 = x * y * 2.0 + imag;
    x = nx;
    y = ny;
    if (x * x + y * y < 4.0) {
        c = c + 1;
    }
}""",
    name=f"mandelbrot",
)
  \end{lstlisting}
\end{figure}

\subsection{Matrix Multiplication}
Matrix multiplication is a fundamental operation in various scientific and technological computations, including deep learning. Here, we show the results of measuring the speed of multiplying two $N \times N$ square matrices in Fig. \ref{fig:matmul}. For $N$=1024, the WebGPU backend achieved 340x speedup over the CPU backend in the Windows+Chrome environment. The WebGPU backend has a faster implementation using shared memory and can compute faster than WebGL on the same hardware and browser. In the Windows environment, we also measured the speed of the native CuPy implementation (without a web browser) and the multi-threaded NumPy implementation. These implementations are highly optimized and it is difficult to obtain comparable speeds on a Web browser. However, WgPy is by far faster than NumPy on the web browser, and in some cases even faster than the native multi-core CPU implementation, which suggests that WgPy can expand the range of tasks that can be implemented on a web application.

\subsection{Training of ResNet-18}
As an example of deep learning application, we train ResNet-18, a type of Convolutional Neural Network (CNN). For the deep learning library, we use Chainer, which can use NumPy and CuPy as backends. Chainer version 5.4.0 is implemented in pure Python, making it possible to run in a web browser with minimal edits, such as removing multiprocess mechanisms. We use CIFAR-100 as the dataset. CIFAR-100 is an image classification dataset consisting of 100 classes, containing 50,000 color images of 32px × 32px. The batch size is set to 16 considering the available memory of mobile devices. The measured speed of training is shown in Fig \ref{fig:resnet}. In the Windows and Chrome settings, the WebGPU backend achieved a 95x speedup over the CPU. Since the most computationally intensive part of ResNet training is the matrix multiplication, high efficiency was obtained with WebGPU, which has more efficient implementation of the matrix multiplication.

\subsection{Hyperparameter Optimization of CNN via Distributed Computing}
With over a billion smartphones in use worldwide, the potential to use these devices for volunteer computing could lead to a vast pool of computational resources. We prototyped a framework that can use smartphones as nodes for distributed computing. The communication part, which needs to be implemented in JavaScript, is included in the framework, so application developers only need to implement the core algorithm in Python to perform distributed computing. An example of a task that can be accelerated through distributed computing is the hyperparameter optimization of neural networks. Sophisticated hyperparameter (the number of layers, the number of channels of each layer, etc.) optimization techniques are studied in the field of neural architecture search (NAS)~\cite{nas2019}. In this experiment, we used the simplest grid search method to train a CNN for each hyperparameter candidate and measured the time to complete the training and evaluation of accuracy on the validation set for all candidates. We used the MNIST dataset, performing 10-class digit image classification using a CNN. A subset of 10,000 images are used for the training and 1,000 images are used for evaluation. The CNN consists of 3 layers, with the number of channels in each layer chosen from \{4, 16, 64, 256\}. For each candidate of the 64($=4^3$) hyperparameters of the model, training was performed for one epoch on the worker device, followed by evaluation, and the accuracy was returned. The iPhone 13 Mini was used as the worker device for distributed computing. A server running Linux was used for job management in distributed computing. The total time to complete training and evaluating of all model candidates are shown in Fig. \ref{fig:nas}. As the number of workers increases, processing speed increases almost linearly. The results demonstrate that hyperparameter search can be accelerated by distributed computing. For more practical use, faster performance can be expected by limiting the models to be evaluated using Bayesian optimization instead of grid search. It is challenging to quantitatively measure the effect of distributed computing because the computational cost varies greatly depending on the results of random sampling. However, we include a sample of NAS using the Optuna hyperparameter tuning library~\cite{optuna2019} in our sample code.

\section{Limitations}
The functionality of NumPy is vast, and implementing all of it requires considerable effort. In this study, we focus on the data types and operations used in deep learning. For example, complex numbers, which are rarely used in deep learning, have not been implemented. Additionally, data types such as 64-bit floating-point numbers, which do not exist in GLSL or WGSL, are virtually impossible to implement.

Since Pyodide is currently the most popular Python interpreter that runs on web browsers, there are some parts implemented using Pyodide's own functions that are not included in the Python language specification. Specifically, they are the identification of the address of the NumPy array in the memory managed by WebAssembly and the mechanism to call JavaScript functions from Python. If another Python implementation becomes more popular in the future, it will be possible to change the implementation of these parts. As long as different Python implementations are based on WebAssembly, similar functionality can be expected to be available.

In deep learning implementations, frameworks like PyTorch~\cite{paszke-2019} and TensorFlow~\cite{abadi-2016}, which have their own unique computation mechanisms different from NumPy, are currently mainstream. Since these tools contain a large amount of native code, porting them to the web browser is not easy. However, if porting to the web browser is attempted, the techniques proposed in this paper could be applied to enable GPU support.

\section{Conclusion}
In this paper, we propose "WgPy," a NumPy-like array library accelerated by GPU, which can be used in a Python interpreter running on a web browser. We implemented an interface that allows the use of WebGL and WebGPU, GPU interfaces available in web browsers, directly from Python. Since code using NumPy is generally implemented assuming synchronous processing, we developed a synchronization mechanism using the Atomics API to enable the use of WgPy with minimal changes to the source code. Additionally, we provided a foundation that allows the implementation of custom kernels, facilitating the acceleration of various operations without requiring in-depth knowledge of WebGL or WebGPU. In our experiments, we successfully executed deep learning using GPUs by combining WgPy with a deep learning framework implemented purely in Python. Furthermore, as an example of distributed computation, we demonstrated an implementation of neural architecture search. While this study offers a means to accelerate various operations through user-defined custom kernels, future work may include developing functionality to automatically integrate complex computations into a single kernel using JIT, thereby providing a means to achieve acceleration with less effort.





\bibliography{main}
\bibliographystyle{mlsys2024}



\end{document}